\documentclass[letterpaper, 10 pt, conference]{Files/ieeeconf}  

\IEEEoverridecommandlockouts                              

\usepackage{subfiles}
\usepackage{hyperref}
\usepackage{mathtools}
\usepackage{commath}
\usepackage[noadjust]{cite}
\newtheorem{definition}{Definition}

\newcommand{\defi}[2]{\begin{definition}[\textbf{#1}] #2 \end{definition}}
\newcommand{\zono}[1]{\left\langle#1\right\rangle}
\newtheorem{theorem}{Theorem}
\newtheorem{assumption}{Assumption}
\newtheorem{remark}{Remark}

\usepackage{algorithm, algorithmic}
\def\@fnsymbol#1{\ensuremath{\ifcase#1\or *\or \dagger\or \ddagger\or
   \mathsection\or \mathparagraph\or \|\or **\or \dagger\dagger
   \or \ddagger\ddagger \else\@ctrerr\fi}}
\newcommand{\ssymbol}[1]{^{\@fnsymbol{#1}}}
\usepackage{amsmath}
\usepackage{amssymb}
\usepackage{booktabs}
\usepackage{graphicx}
\usepackage{xcolor}
\usepackage{subcaption}
\usepackage{comment}
\usepackage{balance}
\title{\LARGE \bf
Data-Driven Reachability Analysis of Pedestrians Using Behavior Modes
}

\author{August Söderlund$^{1,2}$, Frank J. Jiang$^1$, Vandana Narri$^{1,2}$, Amr Alanwar$^3$, and Karl H. Johansson$^1$
\thanks{This work was partially supported by the Wallenberg Artificial Intelligence, Autonomous Systems, and Software Program (WASP) funded by the Knut and Alice Wallenberg Foundation. It was also partially supported by the Swedish Research Council, Swedish Research Council Distinguished Professor Grant 2017-01078 Knut and Alice Wallenberg Foundation Wallenberg Scholar Grant and Swedish Strategic Research Foundation CLAS Grant RIT17-0046.
The work that we introduce in this paper has been carried out at Scania CV AB.}
\thanks{$^1$ Authors are with the Division of Decision and Control Systems, EECS, KTH Royal Institute of Technology, Malvinas väg 10, 10044 Stockholm, Sweden \texttt{\small\{augsod, frankji, narri, kallej\}@kth.se}. They are also affiliated with Digital Futures.}
\thanks{$^2$ The authors are with Research and Development, Scania CV AB, 151 87 Södertälje, Sweden. {\tt\small \{august.soderlund, vandana.narri\}@scania.com.}}
\thanks{$^3$ The author is with the School of Computation, Information and Technology, Technical University of Munich, Heilbronn, Germany and the School of Computer Science and Engineering, Constructor University, Bremen, Germany. {\tt\small alanwar@tum.de.}}
}

\begin{document}
\maketitle
\thispagestyle{empty}
\pagestyle{empty}

\begin{abstract}
In this paper, we present a data-driven approach for safely predicting the future state sets of pedestrians. Previous approaches to predicting the future state sets of pedestrians either do not provide safety guarantees or are overly conservative. Moreover, an additional challenge is the selection or identification of a model that sufficiently captures the motion of pedestrians. To address these issues, this paper introduces the idea of splitting previously collected, historical pedestrian trajectories into different behavior modes for performing data-driven reachability analysis. Through this proposed approach, we are able to use data-driven reachability analysis to capture the future state sets of pedestrians, while being less conservative and still maintaining safety guarantees. Furthermore, this approach is modular and can support different approaches for behavior splitting. To illustrate the efficacy of the approach, we implement our method with a basic behavior-splitting module and evaluate the implementation on an open-source data set of real pedestrian trajectories. In this evaluation, we find that the modal reachable sets are less conservative and more descriptive of the future state sets of the pedestrian.
\end{abstract}

\section{INTRODUCTION}
\label{sec:I}

\begin{figure}[t]
    \centering
    {\setlength{\fboxsep}{0pt}%
    \setlength{\fboxrule}{0pt}%
    \fbox{\parbox{3.3in}{\includegraphics[width=1\linewidth]{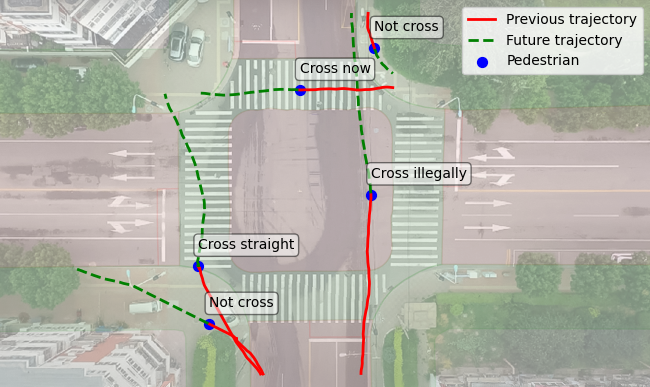}}}}
    \caption{An intersection in Tianjin, China, showcasing the movement of five pedestrians, indicated by the blue circles, and their respective behavior mode, described in the text box above each pedestrian, deduced from either the previous trajectory (red lines) or the future trajectory (dashed green lines).}
    \label{fig:real}
\end{figure}

In urban environments, pedestrians are one of the most vulnerable and challenging road users that automated vehicles need to operate around safely~\cite{schneemann_analyzing_2016}. While pedestrians only have the right-of-way in certain scenarios, they often move around freely, as can be seen in Fig.~\ref{fig:real}, and need to be evaded regardless of how they behave. Moreover, since pedestrian motion is difficult to model accurately, their future states can be hard to predict.

To safely predict the future state sets of other road users, researchers propose using reachability analysis~\cite{althoff_reachability_2010, althoff_set_2021}.
Normally, the reachability of a system is calculated by propagating a known model of the system over a specific time horizon~\cite{althoff_reachability_2010, bansal_hamilton-jacobi_2017}. However, for traffic scenarios involving pedestrians, there is no consensus on how to appropriately model the human and all of its different behaviors~\cite{fan_optimal_2018, meuter_unscented_2008}. Alternatively, using data-driven approaches for the reachability analysis simplifies the problem by removing the need for an explicit model of the system~\cite{alanwar_enhancing_2022, alanwar_data-driven_2021, alanwar_data-driven_2022, devonport_data-driven_2021, chakrabarty_data-driven_2018}. However, data-driven reachability analysis can sometimes lead to predicting overly conservative sets~\cite{alanwar_data-driven_2021, alanwar_data-driven_2022}. One previous work explores the use of known side information to reduce the conservativeness of data-driven reachability analysis~\cite{alanwar_enhancing_2022}. In this work, we will explore the use of behavior predictions to reduce the conservativeness of data-driven reachability analysis.

Several recent works have shown that pedestrians normally communicate their intention in traffic scenarios~\cite{rasouli_understanding_2018, schneemann_analyzing_2016}. For instance, in many scenarios, pedestrians cross the road after achieving eye contact with the driver. 
Also, other external factors that affect their behavior include different aspects in the local scene or the speed of a potential approaching vehicles. To this end, several works aim to model and predict the different behaviors expressed by humans~\cite{lui_modelling_2021, rasouli_multi-modal_2022, peddi_data-driven_2020, rasouli_pedestrian_2020}.
We aim to leverage behavior predictions to
reduce the conservativeness of the data-driven reachable sets by separating the collected pedestrian trajectory data into different modes, representing the different behaviors and intentions of pedestrians in traffic scenarios. 

The main contribution of this paper is providing a modular framework for predicting the future set of states of pedestrians by performing data-driven reachability analysis using the historical trajectories with the same mode as the current behavioral mode of the pedestrian.
The presented approach enables less conservative state predictions in conjunction with an accurate behavior prediction approach while providing similar safety guarantees to the conventional method in~\cite{alanwar_data-driven_2021, alanwar_data-driven_2022}. To the extent of the authors' knowledge, similar research has been conducted in~\cite{hartmann_data-based_2018}; however, our work contrasts in depending on a more computationally efficient set representation called zonotopes. 
Using zonotopes, we can better represent the uncertainties in inputs and noise prevalent in the process, thus rendering guarantees on safety.
More specifically, the contribution of this paper is threefold: 
\begin{enumerate}
    \item We present a modular approach for incorporating behavior mode predictions into data-driven reachability analysis.
    \item We propose an architecture for integrating modal data-driven reachability analysis into automated driving systems.
    \item We implement and evaluate the presented approach on an open-source data set of real pedestrian trajectories. The code to recreate our findings is publicly available\footnote{\href{https://github.com/AugustSoderlund/behavioral-data-driven-reachability}{https://github.com/AugustSoderlund/behavioral-data-driven-reachability}}.
\end{enumerate}

The remainder of the paper is outlined as follows. Section~\ref{sec:II} defines the problem and the necessary prerequisites. Section~\ref{sec:III} introduces the method for the modal data-driven reachability analysis. In Section~\ref{sec:IV}, the setup used for producing the results is introduced. Section~\ref{sec:V} presents the results and discusses the findings. Lastly, in Section~\ref{sec:VI}, the conclusions are summarized along with a short discussion of future work proposals.

\section{PRELIMINARIES}
\label{sec:II} 
In this section, the necessary prerequisites are presented, along with an assumed model of the pedestrian movement system. This section ends with an explicit formulation of the problem that will be solved by using the method presented in Section~\ref{sec:III}.

\subsection{Pedestrian Model}
In this paper, we let the state of a pedestrian be $x = [p_\text{x} \,\, p_\text{y}]^\top \in \mathbb{R}^2$, where $p_\text{x}$ and $p_\text{y}$ are the global x and y positions of the pedestrian, respectively. In this paper, we represent pedestrian dynamics as a linear time-invariant (LTI) discrete model of the form
\begin{equation}
\label{eq:system}
    x(k+1)=Ax(k)+Bu(k)+w(k),
\end{equation}
where $u = [v_\text{x}\,\, v_\text{y}]^\top \in \mathbb{R}^2$ is the x and y components of the pedestrian's velocity and $w\in\mathcal{Z}_w$ is the process noise. Note, that the true pedestrian model $[A \,\, B]$ is considered unknown. We aim to use data-driven reachability analysis to safely approximate this model from historical data.

\subsection{Reachability Analysis and Set Representation}

To represent the future state set prediction of a pedestrian, we use the following definition for reachable sets. 

\defi{Reachable set}{
    The reachable set after $N$ time steps $\mathcal{R}_N$ for the system given in~\eqref{eq:system}, considering the set of inputs $\mathcal{U}_k$ and set of initial states $\mathcal{X}_0$, is defined as
    \begin{align}
        \mathcal{R}_N = \{ x(N) \in \mathbb{R}^{n_x} \bigr\vert &x(k+1)=Ax(k)+Bu(k)+w(k), \nonumber \\ 
        &x(0)\in\mathcal{X}_0, u(k)\in\mathcal{U}_k, w(k)\in\mathcal{Z}_w: \nonumber \\
        &k=0,\dots,N-1 \}
    \end{align}
    We consider a reachable set with $n_x=2$.
}

The reachable set is represented using a zonotope which is defined next. 

\defi{Zonotope \textnormal{\cite{kuhn_rigorously_1998}}}{
\label{def:zonotope}
    Given a center $c_\mathcal{Z} \in \mathbb{R}^{n_x}$ and $\gamma_\mathcal{Z} \in \mathbb{N}$ generators vectors in a generator matrix $G_\mathcal{Z}=\left[ g_\mathcal{Z}^{(1)} \hspace{1.2mm}...\hspace{1.6mm} g_\mathcal{Z}^{(\gamma_\mathcal{Z})}\right]$, a zonotope is defined as 
    \begin{equation*}
        \mathcal{Z} = \left\{  x \in \mathbb{R}^{n_x} \vert x = c_\mathcal{Z} + \sum_{i=1}^{\gamma_\mathcal{Z}} g_\mathcal{Z}^{(i)}\beta^{(i)}_\mathcal{Z}, -1{\leq}\beta^{(i)}_\mathcal{Z}{\leq} 1 \right\}.
    \end{equation*}
We use the shorthand notation $\mathcal{Z}=\langle c_\mathcal{Z}, G_\mathcal{Z} \rangle$ for a zonotope.

    The linear map~\cite{alanwar_distributed_2020} of a zonotope $\mathcal{Z}=\langle c_\mathcal{Z}, G_\mathcal{Z} \rangle$, given $L \in \mathbb{R}^{ n_x^\prime\times n_x}$ is defined and computed as
    \begin{equation}
        L\mathcal{Z} = \left\{ Lz\vert z\in \mathcal{Z} \right\} = \langle  Lc_\mathcal{Z}, LG_\mathcal{Z} \rangle.
    \end{equation}
    
    Given two zonotopes $\mathcal{Z}_1=\langle c_{\mathcal{Z}_1}, G_{\mathcal{Z}_1}\rangle$ and $\mathcal{Z}_2=\langle c_{\mathcal{Z}_2}, G_{\mathcal{Z}_2}\rangle$, the Minkowski sum~\cite{alanwar_distributed_2020} is defined as $\mathcal{Z}_1\oplus\mathcal{Z}_2=\left\{ z_1+z_2\vert z_1\in\mathcal{Z}_1, z_2\in\mathcal{Z}_2\right\}$, and can be computed by
    \begin{equation}
        \mathcal{Z}_1\oplus\mathcal{Z}_2 = \langle c_{\mathcal{Z}_1}+c_{\mathcal{Z}_2}, \left[ G_{\mathcal{Z}_1}, G_{\mathcal{Z}_2}\right] \rangle.
    \end{equation}
    For simplicity, $+$ will be used instead of $\oplus$ to denote the Minkowski sum. Furthermore, we use  $\mathcal{Z}_1 - \mathcal{Z}_2$ to denote $\mathcal{Z}_1 + -1 \mathcal{Z}_2$ not the Minkowski difference.
    
    The Cartesian product between two zonotopes $\mathcal{Z}_1$ and $\mathcal{Z}_2$ is defined and computed as
    \begin{align}
        \mathcal{Z}_1\times\mathcal{Z}_2 = &\left\{ \begin{bmatrix} z_1 \\ z_2 \end{bmatrix} \hspace{0.4mm} \biggr\rvert\hspace{1mm} z_1\in\mathcal{Z}_1, z_2\in\mathcal{Z}_2\right\} \nonumber \\
        & \left\langle \begin{bmatrix} c_{\mathcal{Z}_1} \\ c_{\mathcal{Z}_2} \end{bmatrix}, \begin{bmatrix} G_{\mathcal{Z}_1} & 0 \\ 0 & G_{\mathcal{Z}_2} \end{bmatrix} \right\rangle
    \end{align}
}

\vspace{1mm}
The set of models which is consistent with the noisy historical pedestrians' trajectories is represented using a matrix zonotope which is defined below.

\defi{Matrix zonotope \textnormal{\cite[p.~52]{althoff_reachability_2010}}}{
    A matrix zonotope is defined using a center matrix $C_\mathcal{M}\in\mathbb{R}^{n_x\times p}$ and $\gamma_\mathcal{M}\in\mathbb{N}$ generator matrices $\Tilde{G}_\mathcal{M}=\left[ G_\mathcal{M}^{(1)}\hspace{1.2mm}...\hspace{1.6mm}G_\mathcal{M}^{(\gamma_\mathcal{M})}\right] \in \mathbb{R}^{n_x\times (p\gamma_\mathcal{M})}$ by 
    \begin{equation*}
        \mathcal{M} = \biggr\{ X \in \mathbb{R}^{n_x\times p} \biggr\vert X=C_\mathcal{M} {+} \sum_{i=1}^{\gamma_\mathcal{M}} G_\mathcal{M}^{(i)}\beta^{(i)}_\mathcal{M}, -1{\leq}\beta^{(i)}_\mathcal{M}{\leq} 1\biggr\}.
    \end{equation*}
    Using similar notation as for zonotopes, the shorthand notation is used for a matrix zonotope $\mathcal{M}=\langle C_\mathcal{M}, \Tilde{G}_\mathcal{M} \rangle$.
}

\begin{figure}[b]
    \centering
    {\setlength{\fboxsep}{0pt}%
    \setlength{\fboxrule}{0pt}%
    \fbox{\parbox{3.3in}{\includegraphics[trim={2cm 6cm 7cm 1cm}, clip, width=1\linewidth]{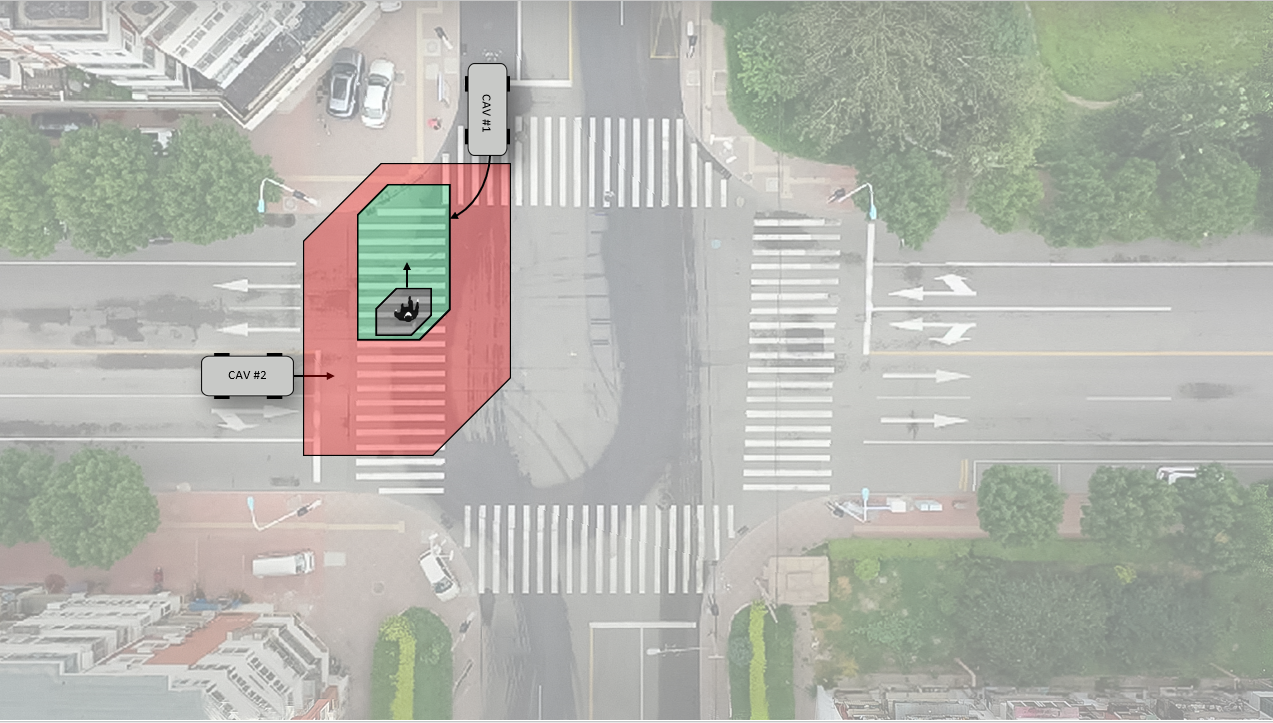}}}}
    \caption{Problem setup with a crossing pedestrian and two approaching connected and autonomous vehicles. The gray set is the vehicles' estimation of the location of the pedestrian. The red set and the green set are the predicted set of states using conventional methods and using proposed techniques in this paper, respectively.}
    \label{fig:problem}
\end{figure}

\subsection{Data-Driven Reachability Analysis}
\label{sec:II.D}
Given $K$ input-state data trajectories of arbitrary lengths $T_i$ for $1\leq i\leq K$, denoted by $\{ u^{(i)}(k)\}_{k=0}^{T_i-1}$ and $\{ x^{(i)}(k)\}_{k=0}^{T_i}$, the input-state matrices essential for data-driven reachability analysis are constructed by
\begin{align}
    \label{eq:traj}
    &X = [  x^{(1)}(0)\dots x^{(1)}(T_1)\dots x^{(K)}(0)\dots x^{(K)}(T_K)], \nonumber \\
    &X_+ = [ x^{(1)}(1)\dots x^{(1)}(T_1)\dots x^{(K)}(1)\dots x^{(K)}(T_K) ], \nonumber \\
    &X_- = [ x^{(1)}(0)\dots x^{(1)}(T_1{-}1)\dots x^{(K)}(0)\dots x^{(K)}(T_K{-}1) ], \nonumber \\
    &U_- = [ u^{(1)}(0)\dots u^{(1)}(T_1{-}1)\dots u^{(K)}(0)\dots u^{(K)}(T_K{-}1) ].
\end{align}
Note that both the first data point and the last data point in each trajectory are dropped in the case of $X_+$ and $X_-$, respectively. The total amount of data points in the shifted matrices is $T=\sum_{i=1}^K T_i$.

The algorithm used for data-driven reachability analysis is presented in~\cite[Algorithm~1]{alanwar_data-driven_2022}. The input-state trajectories are preferably collected offline and stored appropriately for easy extraction and usage. Also, the algorithm yields an over-approximation of the reachable set according to~\cite[Theorem~1]{alanwar_data-driven_2022} given trajectories generated by the unknown system with a structure given in~\eqref{eq:system}. 

\subsection{Problem Statement}

Given the prerequisites previously presented in this section, it is now possible to formulate the problem this paper aims to solve. Our overall goal is to utilize historically collected data on pedestrian trajectories to improve the real-time predictions of pedestrian future state sets, so automated driving systems can navigate with safety guarantees in urban environments. As a motivating example, Fig.~\ref{fig:problem} shows the setup where a pedestrian crosses a zebra crossing with two approaching autonomous vehicles. Conventional methods significantly over-approximate the predictions that hinder both vehicles from proceeding to avoid intersecting with the over-approximated reachable set of the pedestrian. We aim to predict more descriptive reachable sets that enable more efficient planning around pedestrians in urban situations.
 
More specifically, given the current behavior mode of a pedestrian, historically collected pedestrian input-state trajectories $\mathcal{D}=(U_-, X_-, X_+)$, and assumed noise captured by zonotope $\mathcal{Z}_w$, construct the modal input zonotope $\hat{\mathcal{U}}_k$ and compute the modal reachable set $\hat{\mathcal{R}}_N$, which represents the real-time prediction of the future state sets of the pedestrian.

\section{METHODOLOGY}
\label{sec:III}

The proposed method is presented in this section, along with techniques that aim to fully utilize pedestrian behavior in computing the reachable sets.

\subsection{Context: System Architecture}
\begin{figure}[tb]
    \centering
    {\setlength{\fboxsep}{0pt}%
    \setlength{\fboxrule}{0pt}%
    \fbox{\parbox{3.3in}{\includegraphics[width=1.0\linewidth]{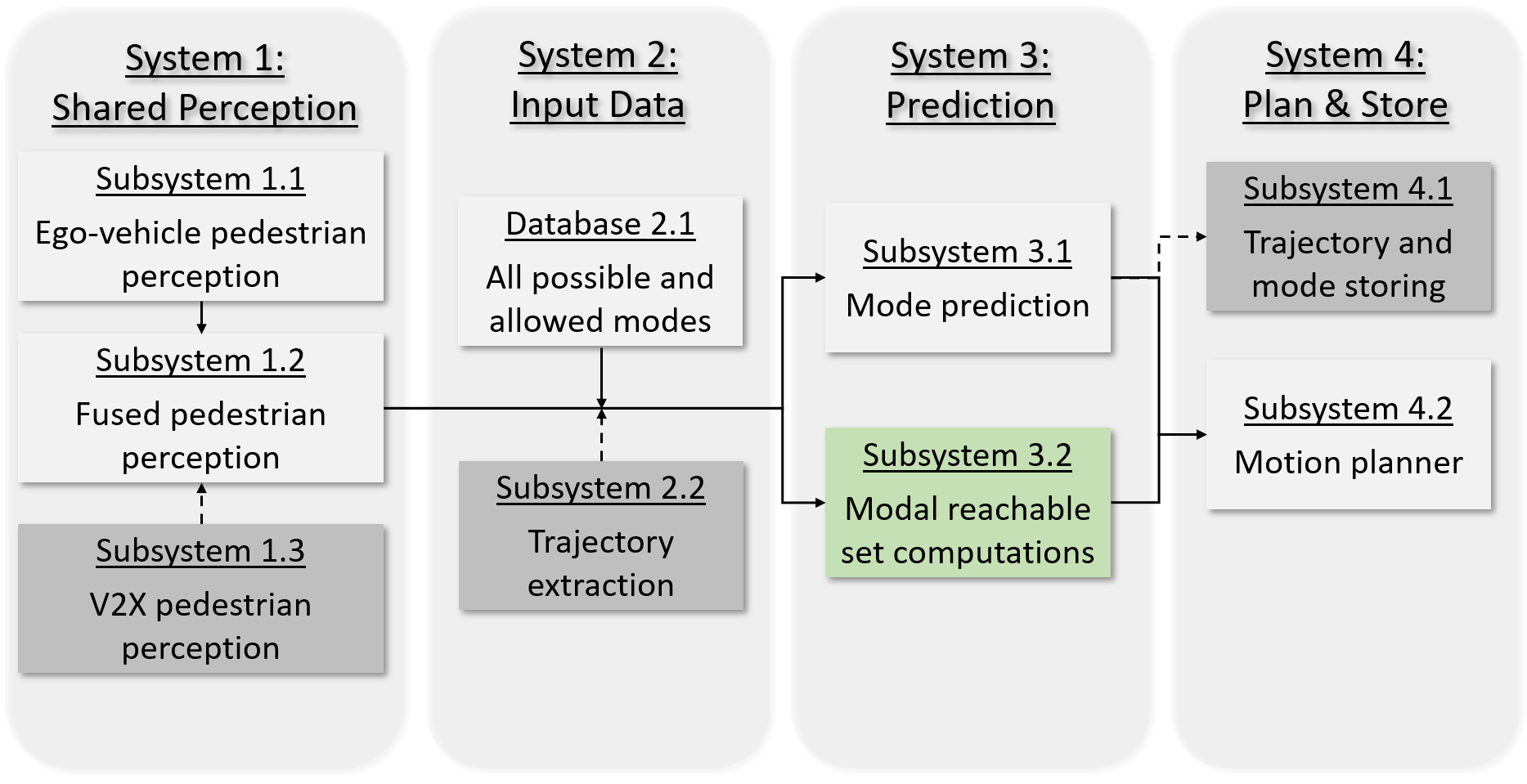}}}}
    \caption{Proposed architecture of systems in a CAV. The light-gray boxes represent internal subsystems, the dark-gray boxes represent external subsystems and the light-green box is the subsystem we focus on in this paper.}
    \label{fig:architecture}
\end{figure}

In this subsection, we contextualize the method proposed in this paper by presenting an architecture illustrated in Fig.~\ref{fig:architecture} for a connected and automated vehicle (CAV) that leverages modal reachable sets.

The architecture starts off with a system that allows
measurements from both internal and external sensors (e.g., information from a camera or LiDAR from another CAV) using sensory data from Subsystems 1.1 and 1.3, as seen in Fig.~\ref{fig:architecture}. 
This information is internally fused in Subsystem 1.2 according to the method presented in~\cite{narri_set-membership_2021}. Then, from database 2.1, we obtain the full set of allowed modes.
To support the modal reachable set computations we extract historical trajectory data from the cloud using the management system in Subsystem 2.2 proposed in~\cite{li_cloud-based_2017}.  The extracted trajectories may be either labeled or unlabeled, since Subsystem 3.2 can perform trajectory labeling before predicting the modal reachable sets. In Subsystem 3.1, we perform the mode prediction of a detected pedestrian, as proposed in~\cite{rasouli_multi-modal_2022}. This mode prediction is used by the motion planner to evaluate how to use the modal reachable sets and not for the modal reachability analysis. In Subsystem 3.2, we compute the modal reachable set computations based on the method presented in this paper. Then, the motion planner in Subsystem 4.2 uses the modal reachable sets as unsafe regions that the CAV is not allowed to traverse.
Finally, this modal trajectory data, together with the mode of the pedestrian, is then sent to the cloud for storage in Subsystem 4.1, using the same cloud-management system previously mentioned.
We leave it as future work on how to retrieve mode prediction certainties in Subsystem 3.1 and how the motion planner should combine these certainties with the modal reachable sets in Subsystem 4.2 to still provide safety guarantees.

For the rest of this section, we will detail the implementation of Subsystem 3.2: the modal reachable set computations.

\subsection{Trajectory Modal Split}
\label{sec:III.B}

To perform modal reachable set computations, we need to take the extracted trajectories from Subsystem 2.2 and split the historical data into chunks of data that we will label with corresponding behavior modes.
This procedure assumes that we have previously collected historical trajectories $\mathcal{D}$ from multiple pedestrians, ideally from a large data set for different road-crossing scenarios.

We start by dividing these trajectories into smaller chunks of length $c_s$, since the behavior along a long trajectory can vary.
We also need to specify a labeling oracle $\mathcal{O}(\mathord{\cdot})$. The intention of this oracle is to take each small chunk of trajectory and label it in accordance with which behavior mode the pedestrian expresses, or is predicted to express. For the sake of evaluation in this work, we implement a simple labeling oracle that has conditional statements on the pedestrian's position and initial heading. Due to our modular approach, the labeling oracle can be easily replaced with other, more advanced implementations.

The result of performing the data division is three lists of length $N_c$: $\mathcal{C}$, $\alpha$ and $\mathcal{L}$, where at index $c$, $\mathcal{C}^{(c)}$, $\alpha^{(c)}$ and $\mathcal{L}^{(c)} = \mathcal O(\mathcal C^{(c)})$ denotes a trajectory chunk, the heading of the pedestrian throughout the chunk, and the behavior label given by the labeling oracle, respectively. We denote the $i$th value in $\mathcal C^{(c)}$ and $\alpha^{(c)}$ with $\mathcal C^{(c)}_i$ and $\alpha^{(c)}_i$.

\subsection{Modal Reachable Set Computation}
Now that the historical trajectories are divided into chunks and labeled, it is possible to perform the modal data-driven reachability analysis, which we summarize in Algorithm~\ref{algo:modal_RA}. Note, Algorithm~\ref{algo:modal_RA} is designed to output the reachable sets for the possible $N_m$ modes found in Database 2.1 in Fig.~\ref{fig:architecture}.

The inputs of the algorithm are the split lists, $\mathcal C$, $\alpha$, and $\mathcal L$, information about the pedestrian's current position and heading, and a heading interval limit $\kappa$ that we will use for finding relevant chunks of trajectory data. For the current position of the pedestrian, we use the estimated state set $\mathcal X_p$ of the locations the pedestrian could be in, which is given to us by the fused perception in Subsystem 1.2 in Fig.~\ref{fig:architecture}. This allows us to start our reachability analysis from a realistic initial condition, where we assume the location of the pedestrian is not known but has to be estimated from sensors. Moreover, we also input the current heading of the pedestrian $\alpha_p$. Then, in lines 2-6, we perform a selection to find the chunks of trajectory data that are relevant for each possible behavior mode in Database 2.1.

\begin{figure}[b]
    \centering
    {\setlength{\fboxsep}{0pt}%
    \setlength{\fboxrule}{0pt}%
    \fbox{\parbox{3.3in}{\includegraphics[trim={0cm 0.7cm 0cm 0.2cm}, clip, width=1\linewidth]{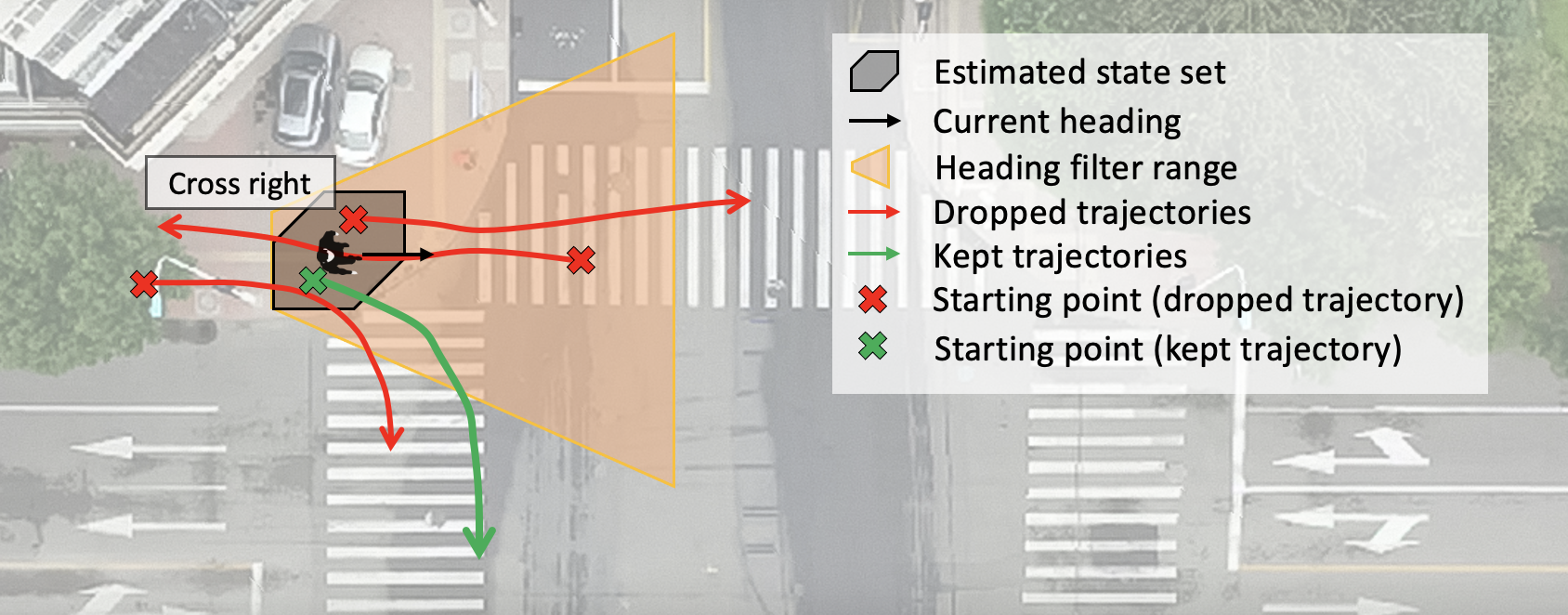}}}}
    \caption{Illustration of selection technique for a particular scenario. The green trajectory is the only trajectory fulfilling the selection criteria where its starting point is contained in the estimated state set of the pedestrian location, its initial heading is within the given heading range, and it has the same mode as the pedestrian. The red trajectories do not fulfill either one criterion or a combination of the criteria, hence not being used in the reachable set predictions.}
    \label{fig:conditional}
\end{figure}

We start by initializing a list for chunks to keep in line 2. Then, in lines 3-6, we iterate through each chunk to check if the chunk is relevant or not. Our criteria for determining if a chunk is relevant or not are:
\begin{itemize}
    \item[(1)] The labeled mode of the chunk must be the same as the possible mode of the detected pedestrian.
    \item[(2)] The starting state $x_c$ of the chunk $\mathcal C^{(c)}_0$ must be inside $\mathcal{X}_p$. 
    \item[(3)] The initial heading $\alpha^{(c)}_0$ must be within the specified heading interval limit $\kappa$, with respect to $\alpha_p$.
\end{itemize}
We illustrate the selection process in Fig.~\ref{fig:conditional}, where only one out of four trajectories is kept.

 \begin{algorithm}[t]
 \caption{Modal Data-Driven Reachability Analysis}
 \label{algo:modal_RA}
 \begin{algorithmic}[1]
 \renewcommand{\algorithmicrequire}{\textbf{Input:}}
 \renewcommand{\algorithmicensure}{\textbf{Output:}}
 \REQUIRE chunks list $\mathcal{C}$, heading list $\alpha$, labels list $\mathcal L$, process noise zonotope $\mathcal Z_w$, estimated set $\mathcal{X}_p$, current heading $\alpha_p$, heading interval limit $\kappa$ 
 \ENSURE  Modal reachable sets $\hat{\mathcal{R}}_k^m, \forall k=1, \ldots, N$ and $\forall m = 1, \ldots, N_m$
  \FOR {$m = 1:N_m$}
    \STATE $\mathcal{C}_\text{keep} = \emptyset$
    \FOR{$c = 0:N_c-1$}
        \IF{the label of the chunk $\mathcal L^{(c)}=m$, starting state $x_c$ of the chunk $\mathcal C^{(c)}_0$ is inside $\mathcal{X}_{k}$, and the initial heading $\alpha^{(c)}_0 \in \left\{\alpha_p+\phi,-\kappa<\phi\leq\kappa\right\}$}
            \item Append $\mathcal C^{(c)}$ to $\mathcal{C}_\text{keep}$
        \ENDIF
    \ENDFOR
      \STATE{$\hat{\mathcal R}_0^m = \mathcal X_p$}
      \STATE{$\hat U_- \gets$ inputs trajectories from $\mathcal C_\text{keep}$}
      \FOR{$k = 0:N-1$}
        \STATE{$\mu_k=\text{mean}\Big(\hat U_-^{(k)}\Big)$}
        \STATE{$\sigma_k = \text{max}\Big(\abs{\hat U_-^{(k)}-\mu_k}\Big)$}
        \STATE $\hat{\mathcal{U}}_k = \zono{\mu_k,\text{diag}(\sigma_k)}$
        \STATE $\hat{\mathcal{R}}^m_{k+1} \gets $ \cite[Algorithm 1]{alanwar_data-driven_2022} with $(\mathcal{C}_\text{keep}, \hat{\mathcal{R}}_{k}^m, \mathcal Z_w, \hat{\mathcal{U}}_k)$
      \ENDFOR
   \ENDFOR
 \end{algorithmic} 
 \end{algorithm}

From lines 7-14, we perform the reachability analysis for the current mode. In lines 7-8, we initialize the reachability analysis. In lines 10-12, we use the input trajectories from the kept chunks to compute the input zonotope from the mean and max deviation in all of the relevant inputs. We compute the input zonotope from the relevant historical data; however, it's also possible to use a predetermined input zonotope for more conservative results, as is done in~\cite{alanwar_data-driven_2022}. Finally, in line 13, the data-driven reachable set for the given mode is computed using~\cite[Algorithm 1]{alanwar_data-driven_2022}.

\begin{assumption}
\label{assumption}
We assume the modal input data $\hat U_-$ contains the bounds of the true, unknown modal input zonotope $\mathcal{U}_k$, $\forall k=0,\dots,N-1$. Thus, we compute the mean and shifted max value of $\hat U_-$, as detailed in line~10-12 of Algorithm~\ref{algo:modal_RA} where the max and mean functions are performed row-wise, and construct $\hat{\mathcal{U}}_k = \zono{\mu_k,\text{diag}(\sigma_k)},$ $\forall k=1,\dots,N-1$. 
\end{assumption}

\theorem{
    Given the modal input-state trajectories $\mathcal{D} = (U_-, X_-, X_+)$ generated by system~\eqref{eq:system} with a full row rank of $\left[ X_-^\top \,\, U_-^\top \right]^\top$, the process noise zonotope $\mathcal Z_w$, and
    assuming that we have a correct mode classifier, a correct labeling oracle, and $\hat{\mathcal{U}}_k, \forall k=1,\dots,N-1$ fulfills Assumption~\ref{assumption}, the reachable set computed from Algorithm~\ref{algo:modal_RA} over-approximates the exact reachable set, hence $\hat{\mathcal{R}}_k \supseteq \mathcal{R}_k$.
}

\begin{proof}
    The exact reachable set, computed using an explicit model of the system, can be established by
    \begin{equation}
        \mathcal{R}_{k+1} = \left[ A_\text{m} \,\, B_\text{m} \right]\left( \mathcal{R}_k \times \mathcal{U}_k \right) + \mathcal{Z}_w,
    \end{equation}
    where $\left[ A_\text{m} \,\, B_\text{m} \right]$ is the true model of the system for a behavior mode. 
    Assuming that the mode of the pedestrian is correctly predicted and that the labeling oracle correctly labels the input-state trajectories, then we know the intention of the pedestrian exactly and the exact actions of pedestrians in previous equivalent conditions. Hence, the behavioral actions observed in the input data are precisely represented in $\hat{\mathcal{U}}_k$, describing the only permissible actions.
    Also, since $\left[ A_\text{m} \,\, B_\text{m} \right]~\in~\mathcal{M}_\Sigma$ according to~\cite[Lemma~1]{alanwar_data-driven_2022} and since both sets $\mathcal{R}_k$ and $\hat{\mathcal{R}}_k$ are computed starting from the same estimated state set $\mathcal{X}_{p}$, it holds that $\mathcal{R}_{k+1}\subseteq \hat{\mathcal{R}}_{k+1}$.
\end{proof}
\begin{remark}
    Bounding $\hat{\mathcal{U}}_k$ to only permit actions prevalent in the data is our method to represent previous behavior. Realistically, a human in traffic scenarios is normally permitted to take actions that are constrained to legal actions (e.g., only walking on sidewalks and crosswalks). However, our implementation constricts the inputs of a pedestrian by prohibiting actions that result in the pedestrian entering regions described by another mode. For example, assuming that the mode classifier accurately predicts that the pedestrian will cross to the right and that historical trajectories are all labeled correctly, we prohibit inputs that result in the reachable set entering the crosswalk straight ahead. As long as these assumptions are satisfied, we can provide safety guarantees. While these assumptions seems restrictive, we believe this enables a principled and less conservative approach to reasoning about the safety of motion around pedestrians.
\end{remark}

\section{EXPERIMENTAL SETUP}
\label{sec:IV}

\begin{table}[b]
\centering
\caption{General parameters}
\label{tab:params}
\begin{tabular}{@{}lccc@{}}
\toprule
\textbf{Description} & \textbf{Symbol} & \textbf{Value} & \textbf{Unit} \\ \midrule
Chunk size & $c_s$ & 90 & [---] \\
Heading interval limit & $\kappa$ & $\pi/4$ & [rad] \\ 
Process noise (for $\mathcal Z_w$) & $\omega$ & $0.005$ & [---] \\
Time horizon & $N$ & $9$ & [s] \\
Estimated set generator matrix & $G_{\mathcal{X}_{k'}}$ & $\begin{bmatrix} 0.5 & 0 & 0.25 \\ 0 & 0.5 & 0.15\end{bmatrix}$ & [m] \\ \bottomrule
\end{tabular}
\end{table}

\begin{figure*} 
\centering 
    \begin{subfigure}[b]{1\textwidth} 
        \centering 
        \begin{subfigure}[b]{0.32\textwidth} 
            \renewcommand\thesubfigure{\alph{subfigure}1} 
            \centering 
            \includegraphics[width=1.0\linewidth]{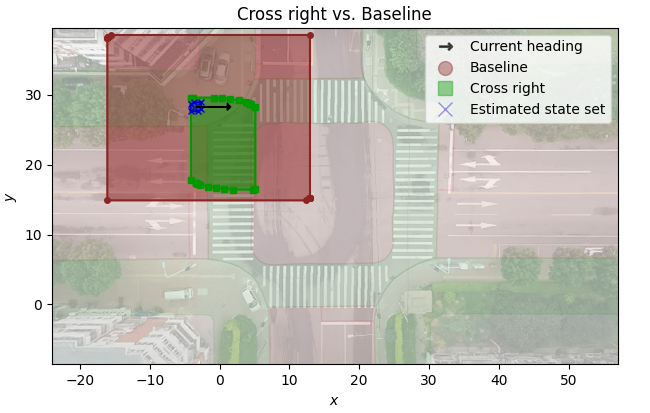} 
            \caption{Cross right} 
            \label{fig:scenarioa1}
        \end{subfigure} 
        \begin{subfigure}[b]{0.32\textwidth} 
            \renewcommand\thesubfigure{\alph{subfigure}1} 
            \centering 
            \includegraphics[width=1.0\linewidth]{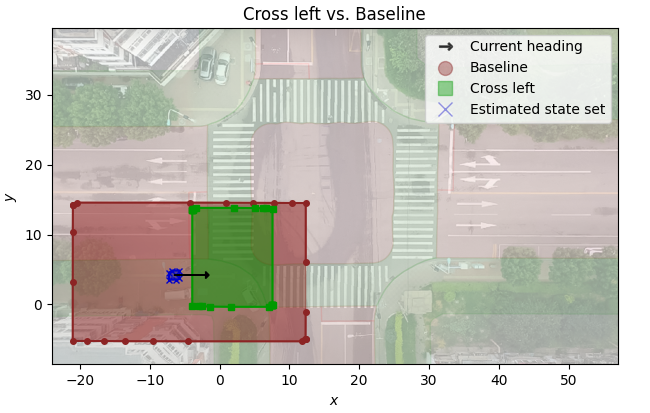} 
            \caption{Cross left} 
            \label{fig:scenariob1} 
        \end{subfigure}
        \begin{subfigure}[b]{0.32\textwidth} 
            \renewcommand\thesubfigure{\alph{subfigure}1} 
            \centering 
            \includegraphics[width=1.0\linewidth]{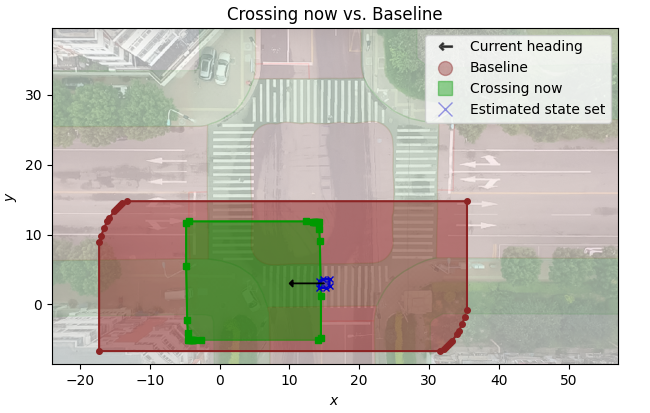} 
            \caption{Crossing now} 
            \label{fig:scenarioc1} 
        \end{subfigure} 
        \addtocounter{subfigure}{-3} 
    \end{subfigure} 
    \begin{subfigure}[b]{1\textwidth} 
        \centering 
        \begin{subfigure}[b]{0.32\textwidth} 
            \renewcommand\thesubfigure{\alph{subfigure}2} 
            \centering 
            \includegraphics[width=1.0\linewidth]{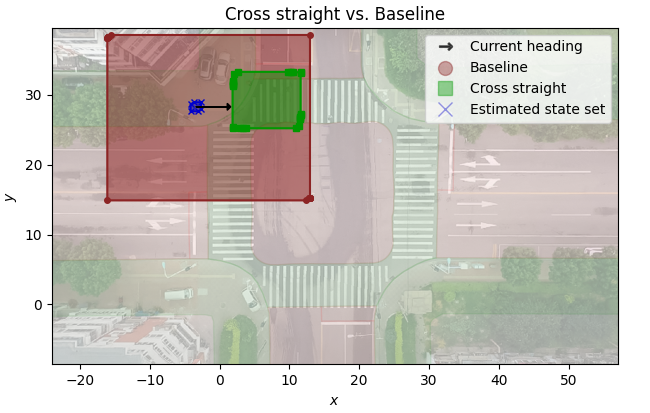} 
            \caption{Cross straight} 
            \label{fig:scenarioa2} 
        \end{subfigure}  
        \begin{subfigure}[b]{0.32\textwidth} 
            \renewcommand\thesubfigure{\alph{subfigure}2} 
            \centering 
            \includegraphics[width=1.0\linewidth]{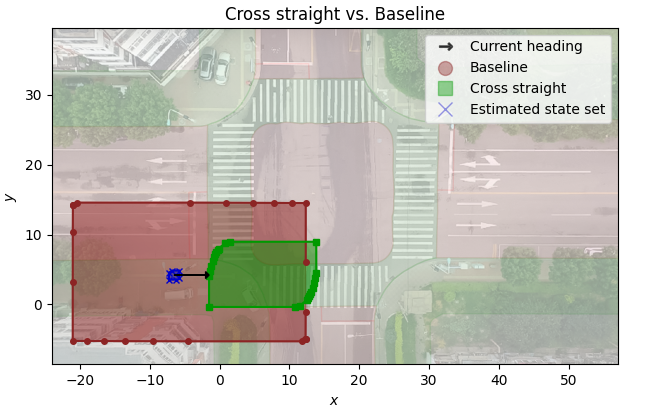} 
            \caption{Cross straight} 
            \label{fig:scenariob2} 
        \end{subfigure} 
        \begin{subfigure}[b]{0.32\textwidth} 
            \renewcommand\thesubfigure{\alph{subfigure}2} 
            \centering 
            \includegraphics[width=1.0\linewidth]{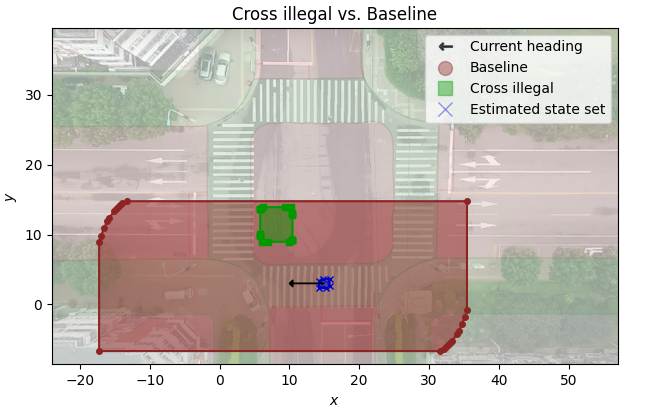} 
            \caption{Cross illegal} 
            \label{fig:scenarioc2} 
        \end{subfigure} 
    \end{subfigure} 
    \caption{ Visualization of the two most descriptive modal reachable sets for Scenario (a), Scenario (b), and Scenario (c).}
    \label{fig:reachability} 
\end{figure*}

\begin{table*}[t]
\centering
\caption{Volume of modal zonotopes for the scenarios (empty fields correspond to infeasible modes for the scenario setup)}
\label{tab:volume}
\resizebox{\textwidth}{!}{\begin{tabular}{@{}l|cccccccc@{}}
\toprule
 & \textbf{Cross left} & \textbf{Cross right} & \textbf{Cross straight} & \textbf{Cross illegal} & \textbf{Cross now} & \textbf{No cross} & \textbf{Unknown} & \textbf{Baseline} \\ \midrule
Scenario (a) & --- & $140.6 \pm 40.72$ & $63.52 \pm 37.04$ & $42.21 \pm 0$ & --- & $93.62 \pm 92.82$ & $18.51 \pm 2.019$ & $652 \pm 247.9$\\
Scenario (b) & $178.8 \pm 62.65$ & --- & $100.3 \pm 45.91$ & $102.2 \pm 77.43$ & --- & $469.2 \pm 142$ & $343 \pm 39.93$ & $867.7 \pm 83.85$\\
Scenario (c) & --- & $80.88 \pm 60.02$ & $217.6 \pm 61.06$ & $750 \pm 2242$ & $414.8 \pm 202.5$ & $151.8 \pm 36.54$ & $185.2 \pm 19.45$ & $1345 \pm 183.8$\\

\bottomrule
\end{tabular}}
\end{table*}

In this section, the setup for producing the results is presented. This includes specifications of parameters, scenario descriptions, and a summary of the data set.

For the evaluating purposes of this paper, the SIND data set~\cite{xu_sind_2022} is used. This data set contains trajectory data for different types of road users (e.g., vehicles, bicycles, pedestrians) from a large signalized intersection in Tianjin, China. We focus on the pedestrians from this data set that consists of their location, velocity, and acceleration for all time steps. 
The data set is also split into training data and testing data. This is done by reserving about $100$ seconds for testing and using the rest of the data set for training.

The considered state-space is $x = (p_\text{x},p_\text{y}) \in \mathbb{R} ^2$, bounded by the spatial limit of the map for the data set, and the input-space is $u=(v_\text{x},v_\text{y}) \in \mathbb{R}^2$. The trajectory data are filtered for velocities below $\|u\|_2<0.5$ m/s, which remove nearly stationary detections beside crosswalks. The modes used in this paper are categorized as crossing events and not crossing events. Specifically, we use \textit{cross left}, \textit{cross right}, \textit{cross straight}, \textit{cross illegal}, \textit{crossing now}, \textit{not crossing} and \textit{unknown}, and other parameters are presented in Table~\ref{tab:params}.

The labeling oracle $\mathcal{O}(\mathord{\cdot})$ forms polygons from given relations in the data set (e.g., crosswalk, road, etc), and a sidewalk is approximated. In this paper, we implement a simple version of a labeling oracle where it uses initial headings of the chunks of trajectories and intersections with the specific polygons and labels the chunk appropriately, according to a set of conditional statements. The crossing modes left, right, straight, and not crossing are all predictions of future events and are labeled when future parts of the chunk intersect legal regions for pedestrians, such as sidewalks and crosswalks. The mode \textit{cross now} is a mode where the pedestrian is currently located inside the crosswalk, and the mode \textit{cross illegal} is both a prediction of a future event and a currently expressed mode and is labeled when the trajectory intersects illegal regions, such as the road. Furthermore, connecting chunks within the same mode along the same full trajectory is also done to remove bias towards areas with a large prevalence of chunks. Also, after the labels have been assigned, some trajectories are dropped to have an even distribution between the different modes.

For comparison, a baseline method is adopted. This baseline also uses the same parameters given in Table~\ref{tab:params}, and it uses Algorithm~\ref{algo:modal_RA} but without forcing the modes of the trajectory chunks and the mode of the pedestrian to be equal, nor do we use a heading filter for the baseline. This enables all chunks that begin inside the initial set to be used in the baseline reachable set predictions.

Specifically, three scenarios will be investigated where the most probable modes can be logically concluded.
\begin{itemize}
    \item[(i)] The initial set is located at the upper-left corner of the sidewalk with an initial heading in the positive x-direction, that is, to the right.
    \item[(ii)] The initial set is located at the lower-left corner of the sidewalk with an initial heading in the positive x-direction.
    \item[(iii)] The initial set is located in the middle of the lower crosswalk with an initial heading parallel to the crosswalk, which is in the negative x-direction.
\end{itemize}
These scenarios will be generated 20 times, and then we compute the mean and standard deviation of the modal zonotope volumes and present it in Table~\ref{tab:volume}, as well as visualizing the modal zonotopes in Fig.~\ref{fig:reachability}. Finally, we run a simulation of the test data where the modal sets and the baseline sets are computed. From this simulation, the state inclusion accuracy is determined. We define the state inclusion accuracy as the portion of times that the true future state of the pedestrian is contained in the predicted reachable set.

\section{EVALUATION}
\label{sec:V}

\begin{figure}[b]
    \centering
    {\setlength{\fboxsep}{0pt}%
    \setlength{\fboxrule}{0pt}%
    \fbox{\parbox{0.5\textwidth}{\includegraphics[width=1.0\linewidth]{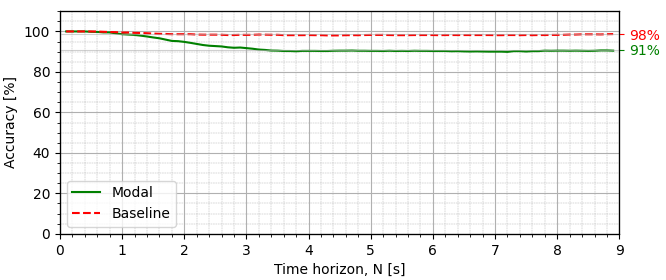}}}}
    \caption{State inclusion accuracy depending on the time horizon.}
    \label{fig:state_inclusion_accuracy}
\end{figure}

In this section, the appropriate results are presented and discussed. In Fig.~\ref{fig:reachability}, we present the modal zonotopes compared to the baseline zonotope for the two most justifiable modes for each scenario. In both analyzed modes in Scenario~(a) (shown in Fig.~\ref{fig:scenarioa1}-~\ref{fig:scenarioa2}), the modal zonotope only intersects parts of the road that it reasonably should (e.g., the modal zonotope for crossing right only intersects the crosswalk that is to the right of the pedestrian). This shows that the state predictions are more descriptive of reality because a pedestrian would reasonably not walk into disallowed areas, such as the middle of the intersection, if they intend to cross the road. In certain driving scenarios for an autonomous vehicle whose starting point is at the top of the intersection, the vehicle would not need to slow down for the pedestrian in Fig.~\ref{fig:scenarioa1} since their modal zonotope does not intersect any path originating from this position, allowing the vehicle to be more efficient regarding brake wear and fuel consumption. This is, on the contrary, untrue when considering the baseline zonotope.

In Scenario~(b), similar advantages for autonomous vehicles with a planned path that ends below the bottom part of the intersection hold for the pedestrian crossing left, visualized in Fig.~\ref{fig:scenariob1}. However, in Fig.~\ref{fig:scenariob2}, the modal zonotope intersects the road in which a pedestrian is not allowed to be located. Upon further inspection of the trajectories used for computing the reachable set, we found that many trajectories originate far down in the initial set, biasing the average velocity across the time steps more homogeneously to the right. This causes the reachable set to propagate into the road. Solving this requires more trajectories in different places across the intersection. Similar reasoning explains the modal zonotope in Fig~\ref{fig:scenarioc1}. Overall, the modal zonotopes are more descriptive than the baseline because they follow the reasonable direction of the mode. 

Based on these findings, we can justifiably establish that the more data available for the specific detection of a pedestrian, the more descriptive and logical the reachable set becomes. Furthermore, given sufficiently many data points, the final reachable set becomes less over-approximate of the predictions, although more importantly, the final reachable set becomes more descriptive of the real outcome. 

We show the mean and standard deviation of the volumes for the zonotopes computed for the scenarios in Table~\ref{tab:volume}. The fields with a solid line indicate modes for the specific scenarios that are illegitimate modes. It means that it is nearly impossible to be within these modes given the surrounding and initial values; thus, consequentially, there are no trajectories within the same mode as the pedestrian. 
We find that all modes visualized in Fig~\ref{fig:reachability} are smaller in volume than the baseline, hence being less conservative. The outlier in Table~\ref{tab:volume} is the mode \textit{cross illegal} in Scenario (b) with an exceptionally large deviation. Upon further investigation, we found that a probable reason for this large deviation is that there are either too few or too different trajectories used in computing the reachable set. However, for all other modes, the modal volume is smaller than the baseline volume, indicating less conservative predictions.

Fig.~\ref{fig:state_inclusion_accuracy} shows the accuracy of the modal and baseline set predictions for varying time horizons. Increasing the time horizon for the predictions lowers the accuracy slightly. Although, the modal accuracy stagnates at around $91$\% for $N>3$ seconds, whereas the baseline accuracy stagnates at around $98$\% for $N>6$ seconds. This shows that it is possible to predict states further in the future using the proposed method with satisfactory accuracy, which would enable more efficient planning for the ego-vehicle. Furthermore, these results use a simplified labeling oracle. Improving the validity of the labeling oracle, in addition to utilizing a mode classifier and combining the modal reachable sets based on the certainties for each mode, would increase the accuracy of the modal reachable sets significantly.

\section{CONCLUSIONS}
\label{sec:VI}

The proposed method of using mode-splitter trajectories for data-driven reachability analysis leads to less conservativeness whilst still providing sufficient safety accuracy. Also, the modular implementation allows for easy modifications to the labeling oracle, improving the results further. In future work, a mode classifier will have to be designed and trained and also reasoned on how to combine the mode confidences with the modal reachable sets. Incorporating temporal logic side information similar to~\cite{alanwar_enhancing_2022} to further reduce conservativeness and include more modes that allow for signalized crosswalks. Furthermore, formulate some metric that better describes the conservativeness and efficiency of the zonotopes relating to the ego-vehicle's planning. Finally, since many intersections are similar, a labeling oracle that generalizes all intersections should be researched.


\balance
\bibliographystyle{ieeetr}
\bibliography{root}

\end{document}